\documentclass[]{spie}  
\usepackage[]{graphicx}
\usepackage{amsmath,amssymb,amsfonts}
\usepackage{algorithmic}
\usepackage{graphicx}
\usepackage{textcomp}

\usepackage{hyperref}

\title{The Efficacy of Microaneurysms Detection With and Without Vessel Segmentation in Color Retinal Images} 

\author{Meysam Tavakoli\supit{a}, Mahdieh Nazar\supit{b}, and Alireza Mehdizadeh\supit{c}
\skiplinehalf
\supit{a}Department of Physics, Indiana University-Purdue University, Indianapolis, IN, USA 46202; \\
\supit{b}Department of Biomedical Sciences, Shahid Beheshti Medical School, Tehran, IRAN; \\
\supit{c}Department of Biomedical physics and Engineering, Shiraz University of Medical Sciences, Shiraz, IRAN
}

\begin{document} 
  \maketitle 

To appear in: Proceedings Volume 11314, Medical Imaging 2020: Computer-Aided Diagnosis; 113143Y (2020) https://doi.org/10.1117/12.2548527
Event: SPIE Medical Imaging, 2020, Houston, Texas, United States

\begin{abstract}
Computer-Aided Diagnosis systems are required to extract suitable information about retina and its changes. In particular, identifying objects of interest such as lesions and anatomical structures from the retinal images is a challenging and iterative process that is doable by image processing approaches. Microaneurysm (MAs) are one set of these changes that caused by diabetic retinopathy (DR). In fact, MAs detection is the main step for identification of DR in the retinal images analysis. The objective of this study is to apply an automated method for detection of MAs and compare the results of detection with and without vessel segmentation and masking either in the normal or abnormal image. The steps for the detection and segmentation are as follows. At the first step, we did preprocessing, by using top-hat transformation. Our main processing was included applying Radon transform, to segment the vessels and masking them. At last, we did MAs detection step using combination of Laplacian-of-Gaussian and Convolutional
Neural Networks. To evaluate the accuracy of our proposed method, we compare the output of our proposed method with the ground truth that collected by ophthalmologists. With vessel segmentation, our algorithm found sensitivity of more than 85\% in detection of MAs with 11 false positive rate per image for 100 color images in a local retinal database and 20 images of a public dataset (DRIVE). Also without vessel segmentation, our automated algorithm finds sensitivity of about 90\% in detection of MAs with 73 false positives per image for all 120 images of two databases. In conclusion, with vessel segmentation we have acceptable sensitivity and specificity, as a necessary step in some diagnostic algorithm for retinal pathology. 
\end{abstract}



\section{INTRODUCTION}
\label{sec:intro}  

Diabetic retinopathy (DR) is a serious eye disease that comes from diabetes mellitus and is the most common cause of vision lost and blindness in the developed countries. Early detection and treatment of DR can avoid people to become affected from this condition or at least the progression of DR can be slowed down~\cite{tavakoli2017automated-onh}. 
Since DR becomes symptomatic at its later stage, patients who deal with the diabetic  may not be informed of having been affected by the DR in its early stage when the treatment is complicated and nearly impossible~\cite{lukac2006color, tavakoli2017comparing}. 
Therefore, mass screening of patients who dealing with diabetes is highly necessary. Consequently, regular retinal experimental tests of the risk groups are mainly advised~\cite{walter2002automatic}. The costs of these tests and lack of experts, especially in rural areas, are the main shortage of this procedure. 
 Moreover, manual screening is slow and resource demanding. 
In the same direction, another challenge is the number of people afflicted with DR which continues to grow at an alarming rate~\cite{thomas2015prevalence, pedro2010prevalence}.  A computer assisted method could assist to deal with these issues. Thus several efforts have been made to establish trustable computer-aided diagnosis (CAD) systems based on color retinal images~\cite{tavakoli2010early,  pourreza2009automatic}. The results indicates that automated DR screening systems are getting closer to be used in clinical settings. The CAD systems are applied for providing physicians assistance at any time and to relieve their workload or iterative works as well, to identify object of interest such as lesions and anatomical structures from the image~\cite{tavakoli2017automated-fovea, tavakoli2017automated-onh}. A critical feature to recognize DR is to detect microaneurysms (MAs) in the automated screening of DR~\cite{tavakoli2013complementary}. MAs are small outpouchings in capillary vessels~\cite{niemeijer2009retinopathy}. They are the first signs of the presence of DR\cite{abramoff2008evaluation}.
In this study we present an effective MA detector based on the combination of preprocessing method, Laplacian-of-Gaussian (LoG) edge detector~\cite{tavakoli2017automated-onh, tavakoli2017effect}, and concept of deep learning~\cite{vedaldi2015matconvnet}. Here, the goal is to compare results of MAs detection with and without vessel segmentation either in normal fundus images or in presence of retinal lesion like in DR.  Before MAs detection in the retinal image, the image has to be preprocessed to ensure adequate level of success in detection. Here we applied  top-hat transform~\cite{tavakoli2017comparing, tavakoli2017effect} to preprocess the images before MAs detection. After preprocessing we used Radon tarnsform edge detector~\cite{tavakoli2013complementary, tavakoli2011radon} for retinal vessel segmentation. After vessel segmentation and masking, a hybrid approach, combining LoG based method and deep learning, was proposed for detection of all the MAs. 

\subsection{Review of Literatures}
There are several approaches for the automatic detection of MAs in color retinal images. 
These studies can be generally categorized into two broad categories unsupervised, and supervised learning~\cite{tavakoli2013complementary, niemeijer2005automatic, dashtbozorg2018retinal, habib2017detection, walter2007automatic, quellec2008optimal, wang2016localizing, zhang2010detection, gegundez2017tool, ram2010successive, lazar2012retinal, mizutani2009automated, dai2018clinical}.
The main benefit of unsupervised approaches is that they do not need a training phase~\cite{habib2017incorporating}. 
Some unsupervised MA detection methods that have been presented are Gaussian filters~\cite{streeter2003microaneurysm, fleming2006automated, niemeijer2005automatic} or their variants~\cite{zhang2010detection, wu2015new}, simple thresholding~\cite{tavakoli2013complementary, giancardo2010microaneurysms}, double ring filter~\cite{mizutani2009automated}, mixture model-based clustering~\cite{sanchez2009mixture} one dimession lines scanning~\cite{lazar2012retinal}, Hessian matrix Eigenvalues~\cite{adal2014automated}, and Frangi filters~\cite{srivastava2015red}.
Different adjustment was applied based on morphology approach to increase the detection accuracy~\cite{fleming2006automated, mizutani2009automated}. Although this type of processing typically is fast and easy to apply, the ability of the approach is limited by the its builder. In better words, some main hidden structures and uncover patterns could be ignored by the builder and cause false segmentation~\cite{dai2018clinical}. 
Different adjustment was used based on morphological function to increase the detection accuracy~\cite{fleming2006automated, mizutani2009automated}. Although this type of processing typically is fast and easy to apply, the ability of the approach is limited by the its builder. In better words, some main hidden structures and uncover patterns could be ignored by the builder and cause false segmentation~\cite{dai2018clinical}. 
Several other mathematical morphology based methods proposed for the detection of red lesions.
According to statistical result, the intensity distribution of MAs is matched to Gaussian distribution~\cite{quellec2008optimal, zhang2010detection, lazar2012retinal}. Therefore, template matching based MA detection approaches were proposed and greatly improved the detection accuracy. 

By growing of machine learning ideas~\cite{tavakoli2019pitching, tavakoli2019bayesian}, studies on MA detection using classification based approaches are mostly seen recently. Antal et al.~\cite{antal2012ensemble} applied a rule-based expert system for MAs detection. In this approach, after selection of MA candidates from retinal images, a rule-based classifier is applied to find true MAs.
Niemeijer et al.~\cite{niemeijer2005automatic} used a hybrid strategy using both top-hat based approach  and a supervised classification. In this approach, MA candidates same as  Antal et al.~\cite{antal2012ensemble} were first selected and then a classifier was trained to differentiate true MAs from false ones. although these machine learning based approaches succeed in detecting hidden structures of features and MAs, they still rely on manually selected features and empirically determined parameters~\cite{dai2018clinical}.
Furthermore, related to third category of detection, learning based approaches,~\cite{gulshan2016development, zhou2017automatic, gargeya2017automated, seoud2015red, abramoff2016improved, orlando2018ensemble, chudzik2018microaneurysm, dai2018clinical, costa2018weakly} proposed to address above issues. Gulshan et al.~\cite{gulshan2016development} presented a deep learning based algorithm to automatically grade DR in retinal images. In this method, deep neural network is applied to process directly the images and output the grading result of DR. While this work successfully addressed the problem of finding hidden structures and empirically determined parameters is not need, for training purpose the classifiers it requires large amounts of retinal images and their annotations, which is costly and time-consuming. Moreover, there is not any  quantitative results generated explicitly for a specific MA. In fact for understanding of the development of DR and monitoring its progress, these quantitative data are critical.
Seoud et al.~\cite{seoud2015red} proposed a novel method for automatic detection of both MAs in color retinal images. The main focus of their work is a new set of shape features, called Dynamic Shape Features, that do not need to precisely segment of red lesion regions. The approach detects all types of MAs  in the images, without distinguishing between them. Differentiating between these lesion types is really important in the clinical practice. The detection of MA is not a practical solution in the medical field. 
Haloi~\cite{haloi2015improved} implemented five layers deep learning with drop out mechanism for diagnosing of early stage DR. The shortcome of the approach was the requirement for a large amount of training data and time-consuming~\cite{wu2017automatic}. 
In comparison, our work focuses on detecting and distinguishing MAs in fundus images. Here, before working on post processing step which we are using the concept of deep learning we add preprocessing unsupervised steps to have some candidates as the MAs and among them using the deep learning we are looking for final true MAs.

\section{Methods} 
\subsection{Databases}

To detect the MAs, two databases (one rural and one publicly available databases) were used. The first rural database was named Mashhad University Medical Science Database (MUMS-DB). The MUMS-DB provided 100 retinal images including 80 images with DR and 20 without DR. The images were obtained 
at 50 degree field of view (FOV) and mostly obtained from the posterior pole view (including ONH and macula) with of resolution $2896 \times 1944$ pixels~\cite{tavakoli2011automated, pourreza2014computationally, tavakoli2011radon}. The second dataset was the DRIVE database consisting of 40 images with image resolution of $768\times 584$ pixels in which 33 cases did not have any sign of DR and 7 ones showed signs of early or mild DR with a 45 degree FOV. For algorithms that operate in a supervised manner this database is often divided into a testing and training set, each containing 20 images. For the test set, two specialists provided manual segmentations for each image. The training set has manual segmentations made by just the first specialist \cite{niemeijer2004comparative}.


\subsection{Preprocessing and Image Enhancement} 
\label{sec:Preprocessing}

\begin{figure}
   \begin{center}
   \begin{tabular}{c}
   \includegraphics[height=6cm]{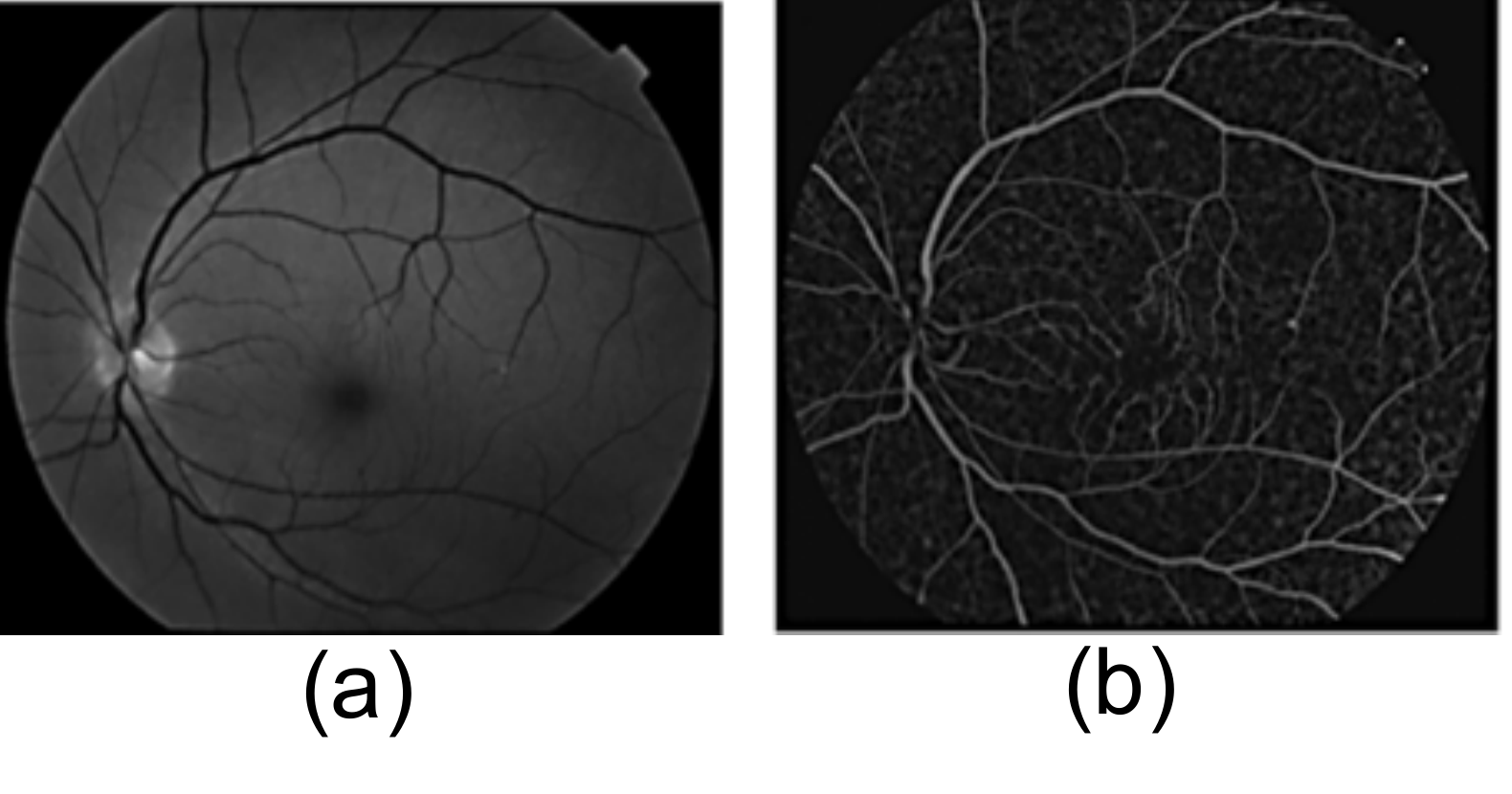}
   \end{tabular}
   \end{center}
   \caption[example] 
   { \label{fig:top-hat} 
The result of Preprocessing step., (a) Green channel image from MUMS-DB (b) Top-hat result}
   \end{figure}

The preprocessing step provides us with an image with high possible vessel, MAs and background contrast and also unifies the histogram of the images~\cite{pourreza2014computationally, tavakoli2017automated-fovea}. Although retinal images have three components (R, G, B), their green channel has the best contrast between vessel and background; so the green channel is selected as input image. The top-hat transform is one of the important morphological operators. In our pre-processing the basic idea is increasing the contrast between the vessels and background regions of the image. A top-hat transformation was based on a disk structure element whose diameter was empirically found that the best compromise between the features and background. The disk diameter depended on the input image resolution. After top-hat transformation, we used contrast stretching to change the contrast or brightness of an image. The result was a linear mapping of a subset of pixel values to the entire range of grays, from the black to the white, producing an image with much higher contrast. The result of first step is shown in Fig.~\ref{fig:top-hat}

\subsection{Vessel Segmentation and Masking} 
The results of preprocessing step were used as input images for main processing. In this study a Radon transform based algorithm was proposed for segmentation of retinal vessels~\cite{tavakoli2013complementary, tavakoli2011radon}. At first, fundus image was partitioned into overlapping widows. To find objects on border of sub-images, we have to define an overlapping pattern of sliding windows. To determine the size of each sub-image or sliding window we used our knowledge database. In this case size of the window was selected two times greater than the thickest vessel. 
The high contrast between blood vessels and image background in preprocessed images caused the vessels to be associated with peaks in the Radon space. The Radon transform, even in the presence of noise, produces peaks from vessel lines in the image domain. The longer the vessel in the image domain, the stronger the maximum peak in the transformed domain will be. (See Fig.~\ref{fig:MA} b,f)

\subsection{Microaneurysm Detection} 
To remove all obstructing effects, we masked our images with the vascular tree mask and remove them from the image.  In this section the LoG edge detector approach following with Convolutional
Neural Networks (CNNs) was used for Extraction of circular pattern not only was utilized for MAs detection also simplified the statistical analysis of the input retinal image. The MAs have a number of features that can be used in their detection: (1) Gaussian shape of the MA cross-sectional grey level profile~\cite{lazar2012retinal}, (2) Their circularity~\cite{tavakoli2013complementary}, and (3) Their size  which is less than $125 \mu m$~\cite{walter2007automatic}. 
The LoG edge detector uses the second-order spatial differentiation. 

The Laplacian is usually combined with smoothing as a precursor to finding edges via zero-crossings. The 2-D Gaussian function:
	\begin{equation}
	\label{eq:Gaussian}
h(x,y)=e^{\frac{-(x^2 + y^2)}{2\sigma^2}},
	\end{equation}
where $\sigma$ is the standard deviation, blurs the image with the degree of blurring being determined by the value of $\sigma$. If an image is pre-smoothed by a Gaussian filter, then we have the LoG operation that is defined: $\bigtriangledown^2 G_{\sigma}\ast I$ where $\bigtriangledown^2 G_{\sigma}(x,y) = \frac{1}{2\pi \sigma^4}(\frac{x^2 + y^2}{\sigma^2} - 2)e^{\frac{-(x^2 + y^2)}{2\sigma^2}}$.

In order to detect qnd segment MAs, circular structures with diameter less than $125 \mu m$~\cite{walter2007automatic} should be excerpted in local sub-images (sub-windows). Therefore, the maximum size of window was chosen twice more than size of the biggest MA. 
The size of the LoG edge detector was chosen equal to maximum diameter of the biggest MA in pixel. Here we found that 18 pixels for MUMS-DB, and 10 pixels for DRIVE database empirically. Moreover, all false MAs such as point noise, end point of vessels, and bifurcations which are similar to MAs (FPs) should be correctly remove from our final extraction. Therefore, to validate MAs, following MAs features in the images were used: Intensity, Size, and Shape. Thresholding is 
an easy solution to the MAs validation problem to compare the peak amplitude intensity with predefined thresholds. Size and shape of candidate were checked.
Unlike unsupervised methods which has no initial labels~\cite{tavakoli2019quantitative} and must find natural clustering patterns in the data, deep learning approach is a learning model that can be applied for classification and regression analysis. Deep learning takes in the clustered bag of features and their corresponding labels (MAs or nonMAs) and determines the predictor clusters for each class. The process is to first train the hidden layers which is done by MA and nonMA candidates. We used CNN implemented in MATLAB~\cite{vedaldi2015matconvnet}. The goal of the CNN classification layer is simply to transform all the net activations to a series of values that can be interpreted as probabilities in the final output layer. To do this, the CNN
MATLAB toolbox is applied onto the net outputs. 

A total number of 140 color retinal images were labeled independently with enough  quality by an expert ophthalmologist with more than 15 years experience in diagnosing DR at early stages. The LoG edge detector produces the MA candidates. These retinal images were made into sub-images, centered on finding the MAs. 1500 sub-images
were applied which mutually agreed with the accuracy
of their clinical label. Among these, 70\% were used
for the training purpose and remaining in the testing set.
Sub-images were extracted at twice the size of the biggest
MA ($125 \mu m$). Normal sub-images (or sub-images without
MA) were taken from the same image as the abnormal
sub-images (or sub-images with MA) in regions that were
free of MA. Overall, 470 sub-images containing MA and
the remaining sub-images without MA were taken from the
whole sub-images.
A single CNN using the MATLAB architecture with a $128 \times 128 \times 3$ input and a four-class output was designed: (1) normal, (2) MAs, (3) bifurcation points, and (4) end points of retinal vessels. The CNN was trained on 1050 sub-images (70\%) and tested on 450 selected sub-images. Training and testing were performed using the MATLAB Deep Learning toolbox~\cite{vedaldi2015matconvnet}.

\begin{figure}
   \begin{center}
   \begin{tabular}{c}
   \includegraphics[height=9cm]{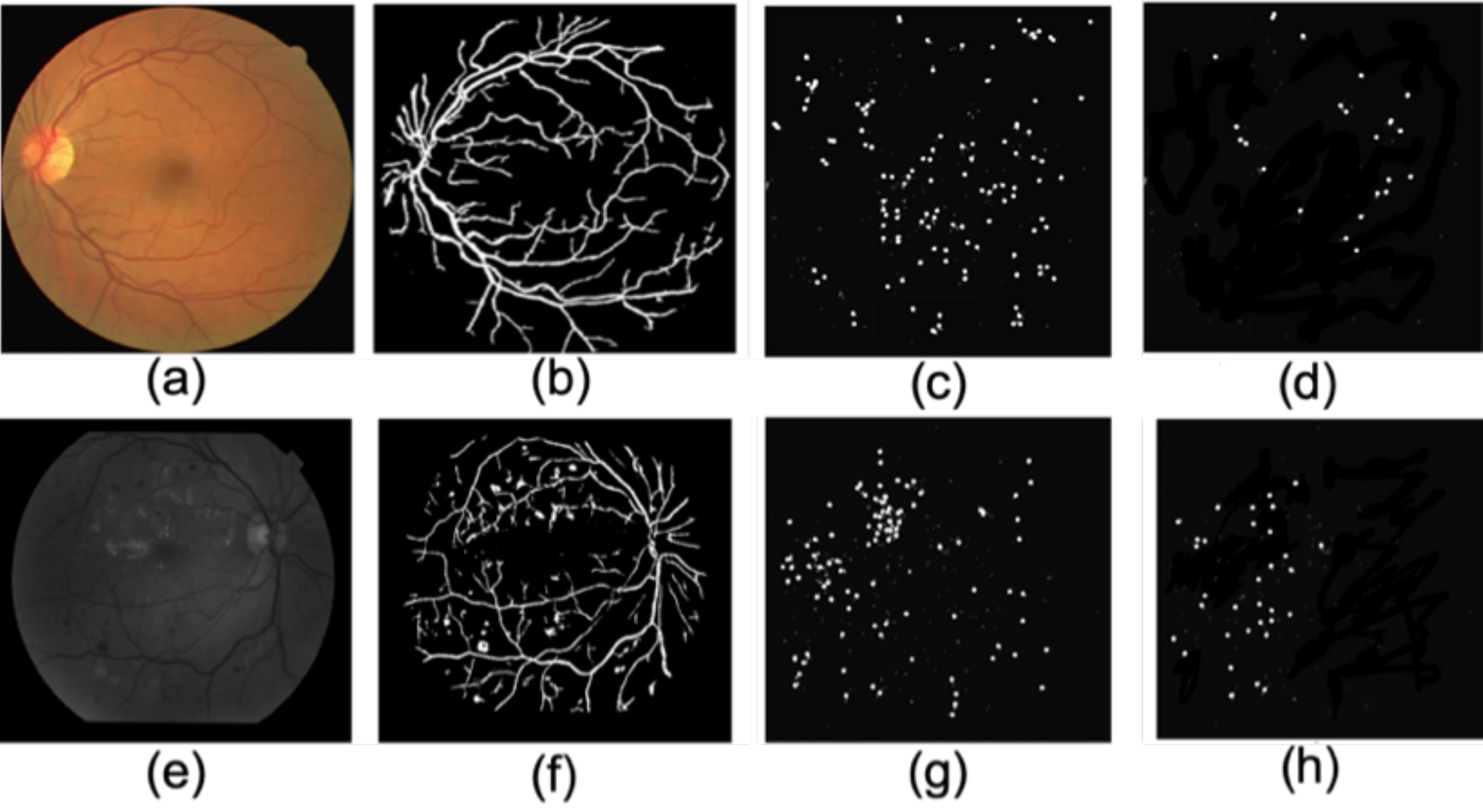}
   \end{tabular}
   \end{center}
   \caption[example] 
   { \label{fig:MA} 
The results for both vessel segmentation and MAs detection in all databases. (a) DRIVE, (b) vessel segmentation of DRIVE image, (c) results of MA detection without vessel segmentation, (d)  results of MA detection with vessel segmentation. (e) MUMS-DB, (f) vessel segmentation of MUMS-DB image, (g) results of MA detection without vessel segmentation, (h)  results of MA detection with vessel segmentation.}
   \end{figure}

As we mentioned, we used preprocessed images until now. Using an overlapping sliding window, the trained CNN was employed over the full scan of the image. A $20 \times 20 \times 3$ window was moved across full sub-images with a slide of 5 pixels overlapping. Each window was the input for a forward pass through our trained CNN, and produce a probability score within that sub-image for each of the four classes of normal and MAs. The result of this sliding window was a blanket of probability values over the entire image for each of the four classes. This procedure took about 1.5 minutes using a PC desktop with an Intel i5-5600HQ Processor. 
The results of MA detection has been shown in the Fig.~\ref{fig:MA}.


\section{Results} 
In this section, statistical information about the sensitivity and specificity measures is extracted. The higher the sensitivity and specificity values, the better the procedure. 
The results for the automated method compared to the ground truth or gold standard were calculated for each image. These metrics are defined as:

\begin{equation} \label{eq:sensitivity}
\begin{aligned}
Sensitivity  &= \frac{TP}{TP+FN} \\
Specificity  &= \frac{TN}{TN+FP} \\
\end{aligned}
\end{equation}

Where TP is true positive, TN is true negative, FP is false positive and FN is false negative same as \cite{tavakoli2013complementary, marin, tavakoli2019quantitative}.

For all retinal images (120 test purpose and  20 training purpose), our reader labeled the both vessels and MAs on the images and the result of this manual segmentation are saved to be analyzed further. According to manual MAs detection by using the Radon vessel segmentation, our automated algorithm finds sensitivity of more than 85\% and specificity 80\% in detection of MAs for 100 color images in MUMS-DB and 20 images of a DRIVE database. Without vessel segmentation, our automated algorithm finds for both sensitivity and specificity of 90\% and 65\% for both in detection of MAs for MUMS-DB and DRIVE databases. Moreover from lesion based view point, With vessel segmentation, our algorithm with above sensitivity had 11 false positives  per image. Also without vessel segmentation, the proposed method finds 73 false positives per image for all 120 images of two databases.

Reaching to the sensitivity of more than 80\% makes our CAD system as good as or better than other related published studies~\cite{kar2017automatic, tamilarasi2015automatic, adal2017automated, quellec2008optimal, zhang2010detection, lazar2012retinal, habib2017detection, wu2015new}. Moreover, this sensitivity in MA based analysis show the ability of our algorithm even in treatment planning and follows up~\cite{orlando2018ensemble}. However, the disadvantage of most of the proposed approaches was that it did not avoid the overfitting issue and was not able to introduce a standard feature selection principle. Besides, for instance, in Ref~\cite{zhang2010detection} the hidden and unnoticeable structures are  ignored. Moreover, a lot of parameters need empirically to be determined.

   \begin{figure}
   \begin{center}
   \begin{tabular}{c}
   \includegraphics[height=5.5cm]{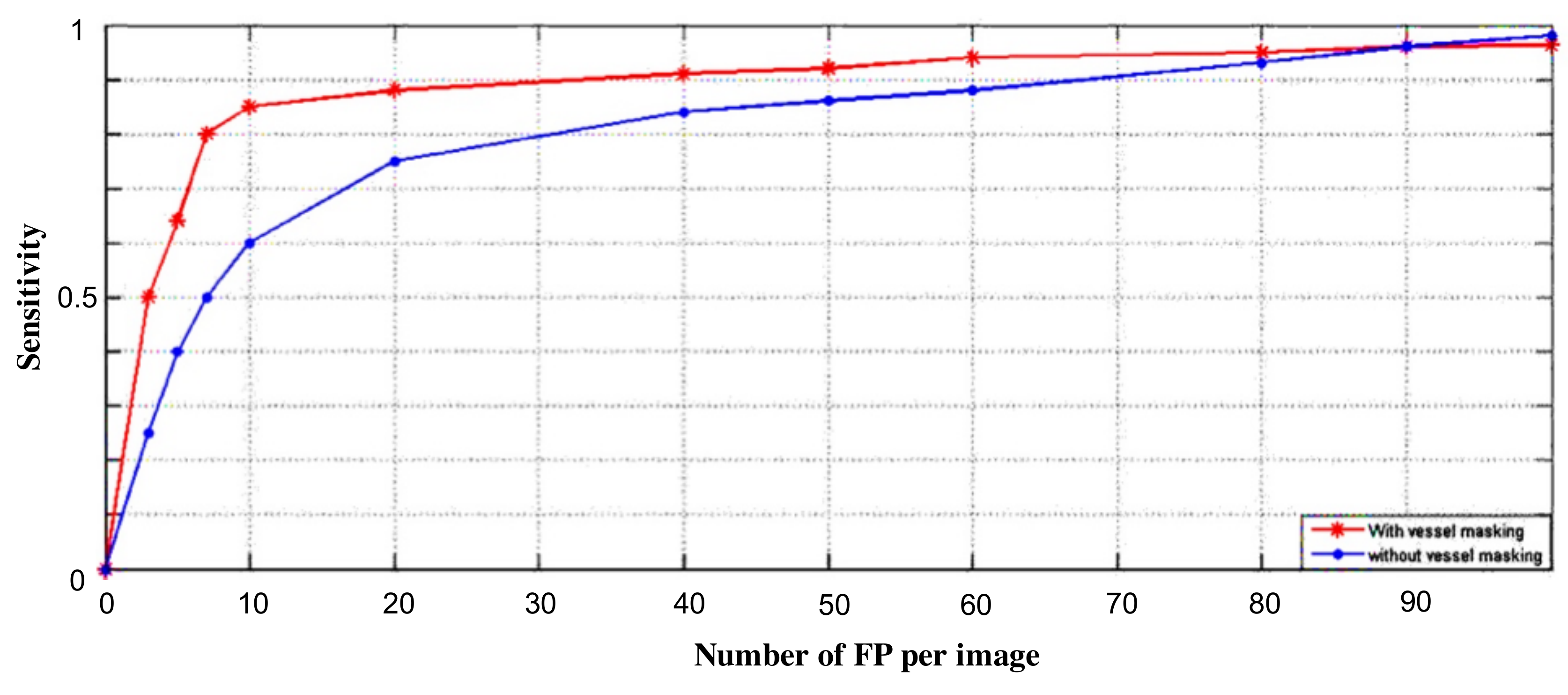}
   \end{tabular}
   \end{center}
   \caption[example] 
   { \label{fig:FROC-MUMS} 
FROC for our method with and without vessel masking in the MUMS-DB.}
   \end{figure} 

   \begin{figure}
   \begin{center}
   \begin{tabular}{c}
   \includegraphics[height=5cm]{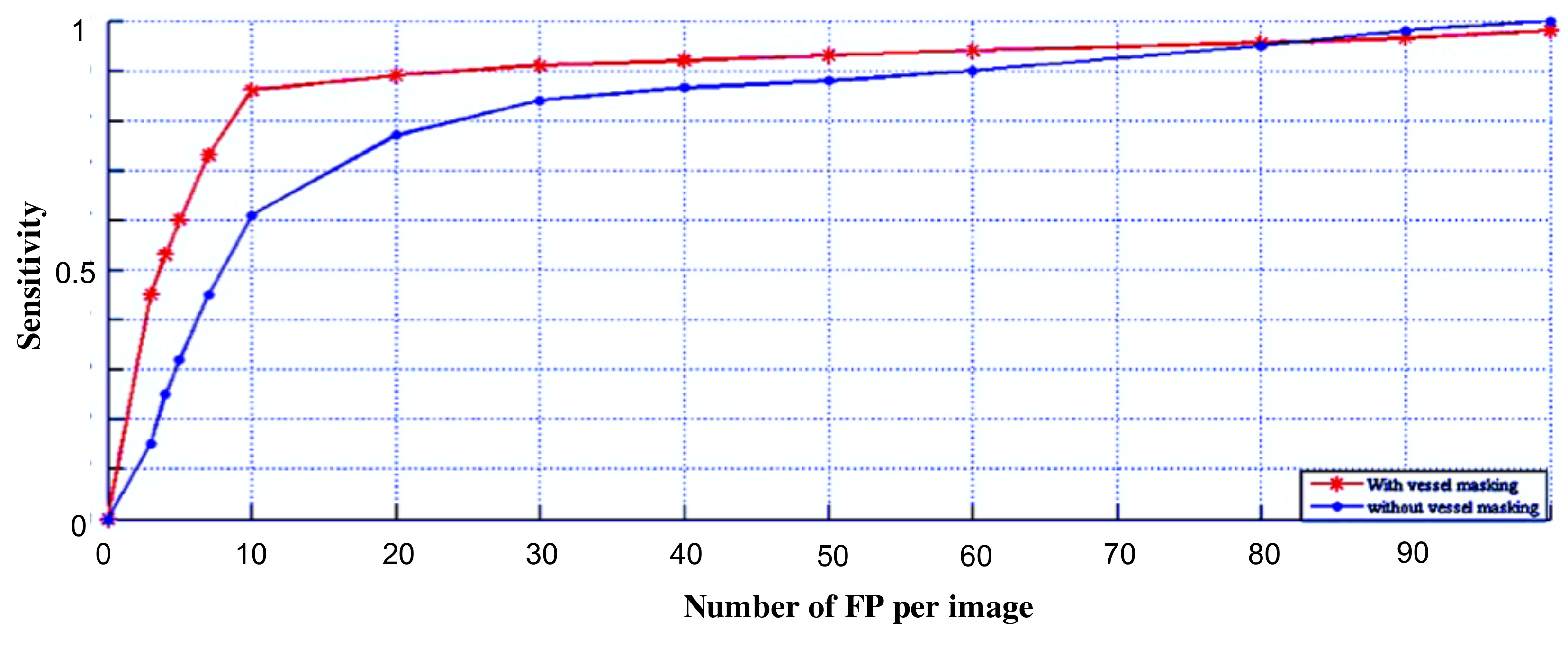}
   \end{tabular}
   \end{center}
   \caption[example] 
   { \label{fig:FROCM-DRIVE} 
FROC for our method with and without vessel masking in the DRIVE dataset.}
   \end{figure} 

\section{Conclusion} 
Although the final purpose of this study was early detection of DR, we were focused only on detection of MAs. The goal of this work was to see the effect of vessel segmentation and masking on final result of MAs detection. 
Since these days medical images are in the digital format, it is feasible to establish a computer-based system that automatically detects landmarks from these images images~\cite{tavakoli2019quantitative, tavakoli2019quantitative-spect}. An automated screening system would save the working load of medical doctors, and letting clinics to use their assets in other important tasks. It could also be doable to check more people and more often by assisting of an automatic screening system, since it would be more inexpensive than screening by humans~\cite{tavakoli2016single}. 
Moreover, computers are suitable to issues involving the derivation of quantitative information from images because of their capacity to process data in fast and efficient manner with a high degree of reproducibility~\cite{matsopoulos1999automatic, tavakoli2017attenuation, welikala2015genetic}.

In this paper, we proposed a hierarchical method based on combination of matching based approach and CNNs to detect all MAs from color fundus retinal image. 
Although the results of MAs detection without vessel masking was promising, to avoid more false positive and screen more patients which they don’t need this is better to segment and mask retinal vessel as a preprocessed task before final MAs detection. Moreover, the results proved that it is possible to use algorithms for assisting an ophthalmologist to segment fundus images into normal parts and lesions, and thus support the ophthalmologist in his or her decision making.
To utilize this program in the follow up of patients, we should add an image registration algorithm so that the ophthalmologist could study the effect of his/her treatment and also the progression of the disease, not only by crisp counting, but also by spatial orientation which is included in presented method. The presented approach was evaluated through a public retinal image database DRIVE. The experiment results demonstrated that using the LoG vessel segmentation and the hierarchical approach has better detection performance in terms of of sensitivity in comparison with other published studies~\cite{tamilarasi2015automatic, adal2017automated, quellec2008optimal, lazar2012retinal, habib2017detection, wu2015new}. 


\bibliography{report}   
\bibliographystyle{spiebib}   

\end{document}